# Experimental demonstration of loop state-preparation-and-measurement tomography

A. F. McCormick,[1] S. J. van Enk,[2] and M. Beck[1,*]

[1]Department of Physics, Whitman College, Walla Walla, Washington 99362, USA

[2]Department of Physics, University of Oregon, Eugene, Oregon 97403, USA

(Dated 1 April 2017)

We have performed an experiment demonstrating that loop state-preparation-and-measurement (SPAM) tomography [C. Jackson and S. J. van Enk, Phys. Rev. A **92**, 042312 (2015)] is capable of detecting correlated errors between the preparation and the measurement of a quantum system. Specifically, we have prepared pure and mixed states of single qubits encoded in the polarization of heralded individual photons. By performing measurements using multiple state preparations and multiple measurement device settings we are able to detect if there are any correlated errors between them, and are also able to determine which state preparations are correlated with which measurements. This is accomplished by going around a "loop" in parameter space, which allows us to check for self-consistency. No assumptions are made concerning either the state preparations or the measurements, other than that the dimensions of the states and the positive-operator-valued measures (POVMs) describing the detector are known. In cases where no correlations are found we are able to perform quantum state tomography of the polarization qubits by using knowledge of the detector POVMs, or quantum detector tomography by using knowledge of the state preparations. Note, however, that the detection of correlated errors does not require estimating any state or measurement parameters.

# I. INTRODUCTION

Quantum mechanics is on the threshold of fundamentally changing modern technology in a number of areas. Commercial quantum cryptographic systems already exist, and are being used for secure communications [1]. High fidelity quantum logic gates have been constructed [2], which is an important step in the construction of a true quantum computer. As these systems improve, better methods for verifying their performance are needed. One such improvement will be minimizing the number of assumptions that are used when characterizing these systems; this will increase our confidence in the reliability of quantum technology.

Quantum tomography is an important tool for characterizing small quantum systems, and presently it comes in several different forms. One form is quantum-state tomography (QST), which estimates the state of a quantum system [3-7]. The system is prepared on many trials, and measurements are performed with a detection system that has many different settings. If enough settings are used the state can be reconstructed. In QST it is assumed that the state is initially unknown, but the detector is completely known and specified.

Another form of tomography is quantum-detector tomography (QDT), which estimates the positive-operator-valued measure (POVM) that describes a detector [8-12]. Here the detector is illuminated with many different probe states, and the detector is characterized by measuring its response to these states. In QDT it is assumed that the states are known and well-characterized, but the detector POVM is initially unknown.

In quantum-process tomography (QPT) the process that transforms an open quantum system from one state to another is fully characterized [13-17]. This is done by having known input states and performing QST on the output states.

There are forms of tomography that lie between QST and QDT. In self-calibrating tomography some state parameters and some detector parameters are assumed known, in order to determine the unknown state and detector parameters [18]. In data-pattern tomography known states are used to calibrate the response of a detector, and this calibration can be incorporated into the state reconstruction procedure [19, 20].

One last form of quantum tomography is state-preparation-and-measurement (SPAM) tomography [21-25]. SPAM tomography attempts to estimate both the state and measurement parameters in a self-consistent manner. Here we are interested in loop (or non-holonomic) SPAM tomography, which involves three separate steps. First, measurements are performed as both the state and detector settings are varied. The Hilbert space dimension is assumed to be known, but neither the states nor the detectors are known, and no other assumptions are made about them. Second, with only this minimal assumption it is possible to determine whether or not there are correlations between the state preparations and the measurements by analyzing the data to look for self-consistency while going around a loop in parameter space, without the need to estimate any state or measurement parameters [23]. Correlated errors are especially detrimental to fault tolerance in quantum computation, so detecting them is important. Finally, if it is determined that there are no correlated SPAM errors, it is then possible to estimate the states using information about the detectors, or vice versa [21, 22].

Here we demonstrate that loop SPAM tomography is capable of detecting correlated errors in the preparation and measurement of qubits encoded in the polarizations of individual photons. Furthermore, we show that it is possible to determine which state preparations are correlated with which measurements.

## II. THEORY

### A. SPAM Correlations

We begin with a brief review of the theory of loop SPAM tomography, based on the discussion in Refs. [23] – [25]. Suppose we have a source that can be prepared in states that are described by density operators $\hat{\rho}_a$, where the subscript labels the different possible state preparations. We also have a detector that is described by the POVM elements $\hat{\Pi}^i$, where the superscript labels the different possible measurements (detector settings). The probability $p_a^i$ of a detection for state preparation $a$ and detector setting $i$ is then given by the Born rule

$$p_a^i = \mathrm{Tr}\left(\hat{\rho}_a \hat{\Pi}^i\right) . \tag{1}$$

State tomography is performed by fixing the (unknown) state preparation $a$, and measuring $p_a^i$ for a number of different detector settings $i$. The measured $p_a^i$'s and the known $\hat{\Pi}^i$'s are then numerically processed (using one of several different techniques [5, 6, 26]) to obtain $\hat{\rho}_a$. In detector tomography the detector setting $i$ is fixed while the state preparation $a$ is varied; measurements of $p_a^i$ and the known $\hat{\rho}_a$'s then determine $\hat{\Pi}^i$. The symmetry between $\hat{\rho}_a$ and $\hat{\Pi}^i$ in Eq. (1) allows us to see how varying one and performing measurements of $p_a^i$ allows the determination of the other.

Equation (1) can be generalized, it need not apply only to POVM elements. It applies equally well to any observable (Hermitian operator) $\hat{\Sigma}^i$

$$S_a^i = \mathrm{Tr}\left(\hat{\rho}_a \hat{\Sigma}^i\right) , \tag{2}$$

where $S_a^i$ is the expectation value of the observable. If one wants to, one can think of the Hermitian operators $\hat{\Sigma}^i$ as a linear combinations of POVM elements.

In real experiments, neither the state preparations $\hat{\rho}_a$, nor the observables $\hat{\Sigma}^i$ can be reproduced with perfect precision. In this case we can still perform measurements, and reconstruct density matrices or observables, but we must more properly consider them as averages over the fluctuations: $\hat{\rho}_a \to \langle\hat{\rho}_a\rangle$, $\hat{\Sigma}^i \to \langle\hat{\Sigma}^i\rangle$, $S_a^i \to \langle S_a^i\rangle$. If there are no correlations between the state preparation and the observables, then there is no problem in using measurements of $\langle S_a^i\rangle$ and a tomographic inversion of Eq. (2) to estimate either $\langle\hat{\rho}_a\rangle$ or $\langle\hat{\Sigma}^i\rangle$. However, if there are correlations then

$$\langle S_a^i\rangle = \langle \mathrm{Tr}(\hat{\rho}_a\hat{\Sigma}^i)\rangle \neq \mathrm{Tr}(\langle\hat{\rho}_a\rangle\langle\hat{\Sigma}^i\rangle) , \qquad (3)$$

so we cannot estimate $\langle\hat{\rho}_a\rangle$ or $\langle\hat{\Sigma}^i\rangle$ individually. The first question that loop SPAM tomography addresses is the detection of such correlated errors, with the only assumption being that the system dimensions are known.

Loop SPAM tomography is useful because new ways to perform tomography [26], and the ability to detect possible tomographic errors are important for quantum information processing applications [17, 22, 27, 28]. For example, new technology is improving the fidelity of quantum logic operations [2]. High precision tomographic measurements are needed to characterize such gates, and experimenters need to be able to place limits on systematic errors in these measurements.

**B. Single-Qubit Loop SPAM Tomography**

In this section we will introduce the notation we use for describing loop SPAM tomography. If there are no SPAM correlated errors, then we can write the expression for the expectation values in Eq. (3) as a matrix equation [23]

$$\bar{S} = \bar{P}\bar{W} . \qquad (4)$$

The overbar indicates a quantity that is expressed as a matrix; it is not an average. The matrix elements for a particular state preparation $a$ and measurement setting $i$ are [25]

$$S_a^i = P_a^\mu W_\mu^i, \tag{5}$$

where summation over repeated upper and lower indices ($\mu = 1, 2, 3$) is assumed, but there is no other distinction made between them. The lower index indicates the matrix row, while the upper index indicates the column. The rows of $\overline{P}$ represent the different state preparations, and the columns of $\overline{W}$ represent the different observables $\hat{\Sigma}^i = W_\mu^i \hat{\sigma}^\mu$. Here $\hat{\sigma}_\mu = \hat{\sigma}^\mu$ plus the identity operator $\hat{1}$ is an operator basis.

In this section we are interested in single qubits. We can represent a general density operator as

$$\hat{\rho}_a = (1/2)\left(P_a^\mu \hat{\sigma}_\mu + \hat{1}\right). \tag{6}$$

We take the operators $\hat{\sigma}_\mu$ to be the Pauli matrices

$$\hat{\sigma}_1 = \begin{pmatrix} 0 & 1 \\ 1 & 0 \end{pmatrix}, \quad \hat{\sigma}_2 = \begin{pmatrix} 0 & -i \\ i & 0 \end{pmatrix}, \quad \hat{\sigma}_3 = \begin{pmatrix} 1 & 0 \\ 0 & -1 \end{pmatrix}, \tag{7}$$

which are orthonormal. This choice is convenient because we are interested in qubits described by the polarizations of individual photons. For such qubits the rows of the matrix $\overline{P}$ are given by the normalized Stokes parameters of the state $\hat{\rho}_a$ as follows: $P_a^1 = s_1$, $P_a^2 = s_2$, $P_a^3 = s_3$ (we are normalizing the Stokes parameters, so we have $s_0 = 1$) [6]. For any given state preparation the three-component vector $P_a^\mu$ falls within a sphere of radius 1, the Poincaré sphere; pure states lie on the surface of the sphere while mixed states are found inside [6, 29].

We can associate the columns of $\bar{W}$ with a detector POVM as follows [23]. Define a two-outcome POVM in terms of elements $\{\hat{E}, \neg\hat{E}\}$ (E and NOT-E). They are written in terms of the matrix elements of $\bar{W}$ as

$$\begin{aligned}\hat{E}^i &= \frac{1}{2}\left[W^i_\mu \hat{\sigma}^\mu + \hat{1}\right] \\ &= \frac{1}{2}\left[W^i_1 \hat{\sigma}_1 + W^i_2 \hat{\sigma}_2 + W^i_3 \hat{\sigma}_3 + \hat{1}\right],\end{aligned} \quad (8)$$

$$\begin{aligned}\neg\hat{E}^i &= \frac{1}{2}\left[-W^i_\mu \hat{\sigma}^\mu + \hat{1}\right] \\ &= \frac{1}{2}\left[-W^i_1 \hat{\sigma}_1 - W^i_2 \hat{\sigma}_2 - W^i_3 \hat{\sigma}_3 + \hat{1}\right],\end{aligned} \quad (9)$$

Here we assume unbiased measurements. For polarization qubits the "direction" of the three-component vector $W^i_\mu$ determines the measurement basis used for detector setting $i$, and $|W^i_\mu|$ determines the detector's discrimination power. Positivity of the POVM is ensured by the inequality $|W^i_\mu| \leq 1$. The two-outcome POVM can be represented in terms of a single observable:

$$\hat{\Sigma}^i = \hat{E}^i - \neg\hat{E}^i = W^i_\mu \hat{\sigma}^\mu. \quad (10)$$

If we examine Eq. (4), we find that the measured expectation values are unchanged under the substitutions

$$\bar{P} \to \bar{P}\bar{G}^{-1}, \quad \bar{W} \to \bar{G}\bar{W}. \quad (11)$$

It can be shown that $\bar{G}$ consists of 9 parameters that are undeterminable by the measurements; these parameters are referred to as blame gauge degrees of freedom [23]. Three of these parameters determine properties such as the choice of Hilbert-space basis, but the others have more interesting interpretations. Despite the fact that these parameters are undeterminable, we will see that it is still possible to detect correlated SPAM errors.

We now turn our attention to detecting correlated SPAM errors. Let $M$ be the number of different state preparations, and $N$ be the number of different detector settings used during the measurements. For the single-qubit case we are considering the state is assumed to be determined by three independent parameters and measurements with $N = 3$ detector settings are sufficient to perform QST. Similarly, the detector POVM is assumed to be determined by three independent parameters and $M = 3$ state preparations are sufficient to perform QDT.

In writing Eq. (4) we assumed that there were no correlated errors between the state preparation and the measurements. However, for $N > 3$ and $M > 3$ the data cannot be completely uncorrelated because there are not enough underlying independent parameters that describe the states and the detectors. Thus, for larger data sets we need to determine when the state preparations and the measurements are effectively uncorrelated, and consistent with the number of independent parameters.

For concreteness, consider the case where $n = 3$ is the number of independent state and detector parameters, and measurements are performed with $M = 2n = 6$ different state preparations and $N = 2n = 6$ detector settings. The 6x6 matrix of expectation values $\bar{S}$ can be partitioned into corners consisting of 3x3 matrices as follows

$$\bar{S} = \begin{pmatrix} \bar{A} & \bar{B} \\ \bar{C} & \bar{D} \end{pmatrix}. \qquad (12)$$

Recall that the rows of $\bar{S}$ refer to a fixed state preparation, while the columns of $\bar{S}$ refer to a fixed detector setting. The matrix elements of $\bar{A}$ are thus determined by state preparations $a = 1, 2, 3$ and detector settings $i = 1, 2, 3$, while the matrix elements of $\bar{B}$ are determined by state preparations $a = 1, 2, 3$ and detector settings $i = 4, 5, 6$. In a similar fashion, the matrices $\bar{C}$ and $\bar{D}$ are determined by different sets of state preparations and detector settings.

Consider matrix $\bar{A}$. This *nxn* matrix consists of enough measurements to be tomographically complete, but because of the undeterminable gauge parameters it is not possible to uniquely determine the states or the detector settings without further assumptions. However, matrix $\bar{A}$ is connected to matrix $\bar{B}$ in the sense that they share a common set of state preparations, and the measured matrix elements of $\bar{B}$ must be consistent with that fact. Matrices $\bar{B}$ and $\bar{D}$ share a common set of detector settings, and their measured matrix elements must be consistent with that fact. Furthermore, $\bar{C}$ and $\bar{D}$ must be consistent with a common set of state preparations, and $\bar{A}$ and $\bar{C}$ must be consistent with a common set of measurement settings.

Another way to look at the connectedness of the corner matrices in Eq. (12) is as follows. If one knew the detector POVM's for settings $i = 1, 2, 3$ used to measure the matrix elements of $\bar{A}$, one could perform QST to estimate the three state preparations $a = 1, 2, 3$. One could then use these known states to perform QDT on the data in $\bar{B}$ to estimate the detector POVM's for measurement settings $i = 4, 5, 6$. These detector POVMs could be used to perform QST on the data in $\bar{D}$ to estimate the state preparations $a = 4, 5, 6$. Finally, these states and the original known detector POVM's for settings $i = 1, 2, 3$ must be consistent with the data in $\bar{C}$. This is what we mean by looking for self-consistency while going around a loop in parameter space.

Define the partial determinant of $\bar{S}$ as [30]

$$\Delta(\bar{S}) \equiv \bar{A}^{-1}\bar{B}\bar{D}^{-1}\bar{C} . \tag{13}$$

Jackson and van Enk have shown that the measured data are internally consistent as described above, and free of correlated SPAM errors under the condition that

$$\Delta(\bar{S}) = \bar{1}, \tag{14}$$

where $\bar{1}$ is the 3x3 identity matrix [23].

Thus, the procedure to detect correlated SPAM errors for a single qubit is as follows. Measure expectation values for $M = 2n = 6$ different state preparations and $N = 2n = 6$ different detector settings (36 total measurements). Construct the matrix of these expectation values $\bar{S}$ as given in Eqs. (5) & (12), and then calculate the partial determinant $\Delta(\bar{S})$ given in Eq. (13). If $\Delta(\bar{S}) - \bar{1} = 0$, to within the statistical errors of the measurements, there is no evidence for correlated SPAM errors. Note that this determination is made by knowing the dimension of the Hilbert space, but is independent of any other assumptions about the state preparations or the measurements. Furthermore, we do not need to estimate any of the parameters that describe the states or the measurements in order to detect the presence of correlated errors. Once it is determined that there are no correlated SPAM errors, it is then possible to use information about the detector settings in order to estimate the states, or vice versa, using standard QST or QDT.

Finally, it is possible to reduce the number of needed state preparations and measurement settings from $2n = 6$ to $n + 1 = 4$, which reduces the total number of measurements from 36 to 16. For example, consider the matrix $\bar{B}$, which has the same state preparations as $\bar{A}$. We need at least one of the detector settings that determine $\bar{B}$ to be different from those that make up $\bar{A}$, but we don't need to change all of the detector settings. Thus, for example, the first column of $\bar{B}$ could use a different detector setting than the first column of $\bar{A}$, but the data in the other columns of $\bar{B}$ can simply be duplicates of the other columns of $\bar{A}$.

Thus, $\bar{S}$ is a $2n \times 2n$ matrix, but it can be constructed from $(n+1)^2$ measured expectation values as follows. The first $(n+1)$ rows and $(n+1)$ columns are made up of the independent elements. Columns $(n+2)$ through $2n$ are copies of columns 2 through $n$, and rows $(n+2)$ through $2n$ are copies of rows 2 through $n$.

## III. EXPERIMENTS

### A. The Experimental Apparatus

In our experiments we use a 150 mW, 405 nm laser diode to pump a 3 mm long beta-barium borate (BBO) crystal. This produces type-I spontaneous parametric downconversion at 810 nm, with signal and idler beams making angles of 3° from the pump. The idler beam is focused into a single-mode, polarization-preserving optical fiber, filtered by a 10 nm bandpass filter centered at 810 nm, and detected by a single photon counting module (SPCM). Detection of an idler photon heralds the production of a single photon in the signal beam. For heralding we use a coincidence unit based on a Xilinx SP605 development board that has a coincidence window of 2.5 ns [31]. The signal beam is filtered with RG780 colored glass and focused into a single-mode, polarization-preserving optical fiber. The output of this fiber is then collimated, and emerges as the "Source" in Fig. 1. Our source has a heralding efficiency of approximately 13%, a heralded single-photon production rate of ~9,000 s$^{-1}$, and a degree of second-order coherence $g^{(2)}(0) = 0.024 \pm 0.003$.

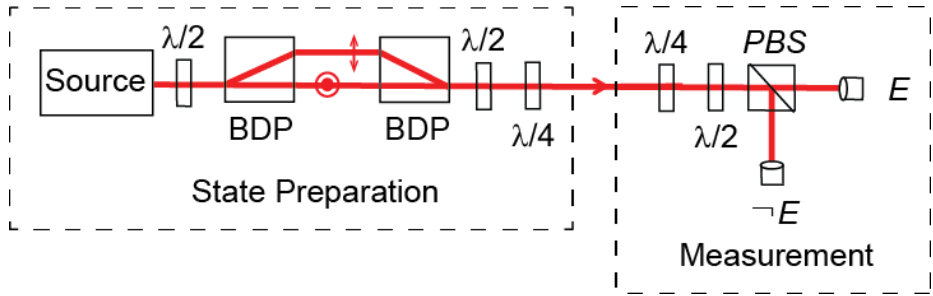

FIG. 1. Experimental configuration for performing single-qubit SPAM tomography. The source produces heralded, linearly-polarized single photons. Here λ/2 denotes a half-wave plate, λ/4 denotes a quarter-wave plate, BDP

denotes a beam-displacing polarizer, PBS denotes a polarizing beam splitter, and $E$ and $\neg E$ are single-photon-counting modules.

The experimental arrangement for loop SPAM tomography is shown in Fig. 1. Linearly polarized photons from the source pass through a half-wave plate that rotates their polarization. These photons then pass through a beam-displacing polarizer (BDP) that spatially displaces the horizontal component of the polarization from the vertical component; the fraction of the horizontal and vertical components is adjusted by the rotation angle of the half-wave plate. A second BPD spatially recombines the beams, but the horizontal component is delayed by a time longer than the coherence time of the individual photons. Thus, the quantum state after the second BDP is a mixture of horizontal and vertically polarized photons. We perform measurements for two different possibilities: a horizontally polarized pure state $\hat{\rho}_H = |H\rangle\langle H|$ (we block the vertically polarized beam to improve the purity), and a mixed state $\hat{\rho}_M = (3/4)|H\rangle\langle H| + (1/4)|V\rangle\langle V|$.

In order to construct the matrix $\bar{S}$ in a given experiment, the state emerging from the BDPs is fixed to be either $\hat{\rho}_H$ or $\hat{\rho}_M$, and the state for preparation $a$ is determined by the rotation angles of the half- and quarter-wave plates that immediately follow the second BDP (Fig. 1). The detector POVM for a given setting $i$ is determined by the rotation angles of the quarter- and half-wave plates that immediately precede the polarizing beam splitter (PBS). The settings are chosen to sample the Hilbert spaces of both the states and the detector POVMs; for both the state preparations and the measurements the quarter-wave plate rotation angles are $\{0, \pi/4, \pi/4, \pi/16, 5\pi/16, 5\pi/16\}$ and the half-wave plate angles are $\{0, 0, \pi/8, \pi/16, \pi/16, 3\pi/16\}$. We have performed experiments using both $2n$ and $n+1$ state and measurement settings; when using $n+1$ settings we use the first four wave-plate angles listed above. In order to calculate statistical errors we perform 10 sequential measurements of $\bar{S}$, and we quote our errors as standard deviations of these measurements.

**B. No Correlated Errors**

First we performed measurements with the source in the pure state $\hat{\rho}_H$ using $n+1$ state and measurement settings. In these measurements the states and measurements were, to the best of our knowledge, independent of each other, so there should be no correlated SPAM errors. Figure 2 shows the results for the mean of $\Delta(\bar{S}) - \bar{1}$, the standard deviation of $\Delta(\bar{S}) - \bar{1}$, and the ratio of these two quantities. By examining the absolute value of the ratio of the mean to the standard deviation [Fig. 2(c)] we find that all of the matrix elements of $\Delta(\bar{S}) - \bar{1}$ are 0 to within half of a standard deviation, which indicates that no correlated SPAM errors were detected.

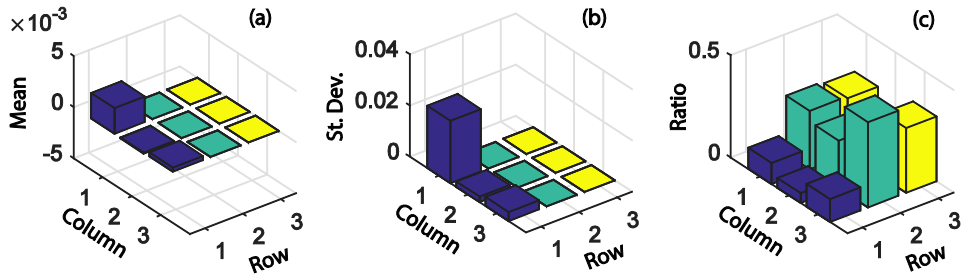

Fig. 2. For no SPAM correlations we show, (a) the mean of $\Delta(\bar{S}) - \bar{1}$, (b) the standard deviation of $\Delta(\bar{S}) - \bar{1}$, and (c) the absolute value of the ratio of these two quantities (mean divided by standard deviation).

Other than assuming that the Hilbert space dimension is known, no assumptions about the state preparation or detector observables were made when processing the measured expectation values to obtain the results shown in Fig. 2. However, since no correlated SPAM errors were found, we should be able to estimate the states and the detector POVMs by now making some assumptions. A very simple way to do this is to note that we can solve Eq. (4) for the desired quantity. For example, if we know the detector POVMs, which are described by $\overline{W}$, and use the measured $\overline{S}$, we can perform QST and reconstruct the matrix that determines the state preparations $\overline{P}$:

$$\overline{P} = \overline{S}\overline{W}^{-1} . \tag{15}$$

Similarly, if we instead know the state preparations we can perform QDT and estimate $\overline{W}$, which specifies the detector observables:

$$\overline{W} = \overline{P}^{-1}\overline{S} . \tag{16}$$

This technique is not guaranteed to produce physically real states or POVMs, as the magnitudes of the three-component vectors that specify the states and POVMs must satisfy $|P_a^\mu| \leq 1$ and $|W_\mu^i| \leq 1$. To ensure that our states and POVMs are physical, if $|P_a^\mu|$ or $|W_\mu^i|$ are found to be larger than 1, we renormalize them so that they are equal to 1 [26, 32]. Our purpose here is not necessarily to perform the most accurate tomography using our measured data (there are other techniques capable of performing more accurate tomography [6, 26]). Rather, we mean to show that this simple technique of matrix inversion can yield a very reasonable estimate of the measured states and POVMs, once we have determined that there are no SPAM correlations.

We don't need to know all of the detector settings or state preparations. For example, knowing the detector POVMs for the first three settings, $i = 1, 2, 3$ is sufficient to estimate all six of the state preparations and the three remaining detector POVMs by using the technique described in Sec II.B [in the paragraph prior to Eq. (13)]. We have used this technique to estimate the reconstructed density operators $\hat{\rho}_a^{(rec)}$ for our measured data. To compare the theoretically expected state preparation $\hat{\rho}_a^{(th)}$ (as predicted from the experimental parameters) to the reconstructed state $\hat{\rho}_a^{(rec)}$ we use the fidelity $F$, given by [6, 33]

$$F_a = \left( \mathrm{Tr}\left[ \sqrt{ \sqrt{\hat{\rho}_a^{(th)}} \hat{\rho}_a^{(rec)} \sqrt{\hat{\rho}_a^{(th)}} } \right] \right)^2 . \tag{17}$$

The fidelity takes on values $0 \leq F \leq 1$, with $F = 1$ corresponding to $\hat{\rho}_a^{(th)} = \hat{\rho}_a^{(rec)}$. Using the same data used to produce Fig. 2, we find that the fidelities of the reconstructed states are all greater than 0.99.

We also reconstruct the POVM elements $\hat{E}^{i\,(rec)}$ [Eq. (8)]. To compare $\hat{E}^{i\,(rec)}$ to the theoretically expected observables $\hat{E}^{i\,(th)}$ we can also use the fidelity (with the density operators replaced by the POVMs). We find that the fidelities are all greater than 0.99. Another way to compare $\hat{E}^{i\,(rec)}$ to $\hat{E}^{i\,(th)}$ is to use the relative error [34]

$$RE^i = \frac{\left\| \hat{E}^{i\,(rec)} - \hat{E}^{i\,(th)} \right\|_2}{\left\| \hat{E}^{i\,(th)} \right\|_2} , \tag{18}$$

where $\|\hat{O}\|_2 = \sqrt{\mathrm{Tr}(\hat{O}^\dagger \hat{O})}$ is the Frobenius norm. We find that the relative errors are all less than 0.023.

We have also performed measurements in which the state after the BDPs is the mixed state $\hat{\rho}_M$. Again we find that $\Delta(\bar{S}) - \bar{1}$ differs from 0 by less than half a standard deviation, and the fidelities and relative errors of the reconstructed states and POVMs are essentially the same as those described above for the pure state $\hat{\rho}_H$.

Finally, we have performed experiments using $2n$ state and measurement settings, for both pure and mixed state preparations. The results are similar to those described above, the only difference being that the fidelities are slightly lower ($> 0.97$) and the relative errors are slightly larger ($< 0.060$). With $2n$ settings we reconstruct 6 different states and POVMs, while for $n+1$ settings we reconstruct only 4. Furthermore, $2n$ settings require more than twice as many measurements as $n+1$ settings (36 versus 16). Both of these factors allow experimental imperfections more opportunities to influence the results.

**C. Correlated Errors**

We have performed an experiment where for state preparation $a=1$ detector setting $i=1$ is modified; this changes the single expectation value $S_1^1$ from what it would be if there were no SPAM correlations. We do this by rotating the detector half-wave plate by $\pi/4$ from where it should be. Everything else is the same as described above, and again we begin with photons in the pure state $\hat{\rho}_H$ and use $n+1$ settings. Without the correlated error we would expect $S_1^1$ to be 1, while with the error it is expected to be $-1$; we measure $S_1^1 = -0.9971 \pm 0.0009$. Figure 3(a)-(c) shows the results for the mean of $\Delta(\bar{S}) - \bar{1}$, the standard deviation of $\Delta(\bar{S}) - \bar{1}$, and the absolute value of the ratio of these two quantities when this correlated SPAM error is present. In this case there is clearly a statistically significant difference of one of the matrix elements of $\Delta(\bar{S}) - \bar{1}$ from 0; it differs from 0 by 48 standard deviations. This indicates that we are able to detect this correlated SPAM error, without having to estimate any of the state or measurement parameters.

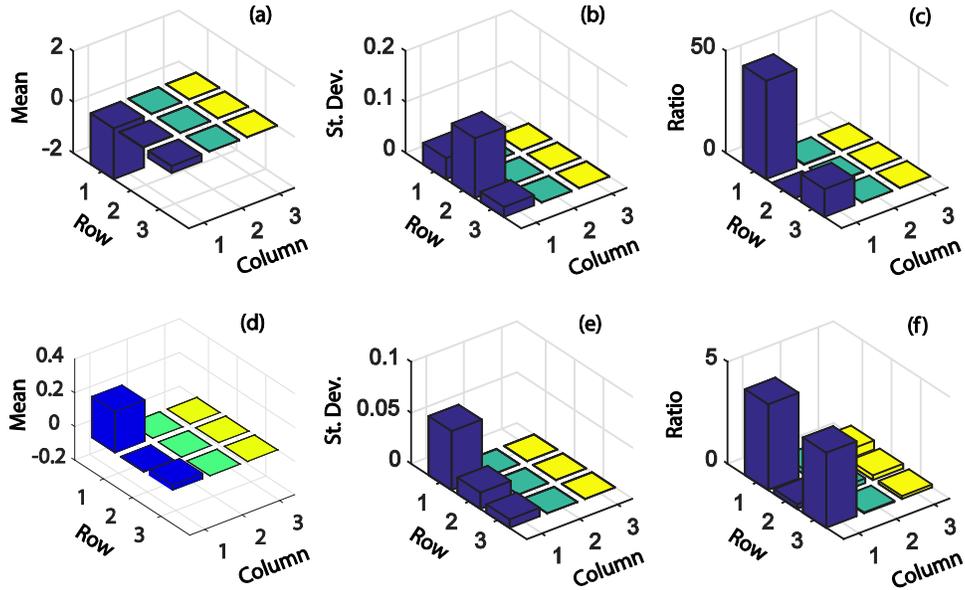

Fig. 3. For measurements with SPAM correlations in $S_1^1$ we show, (a) and (d) the mean of $\Delta(\bar{S}) - \bar{1}$, (b) and (e) the standard deviation of $\Delta(\bar{S}) - \bar{1}$, and (c) and (f) the absolute value of the ratio of these two quantities (mean divided by standard deviation). In (a)-(c) $S_1^1 = -0.9971 \pm 0.0009$, while in (d)-(f) $S_1^1 = 0.79 \pm 0.04$.

In the above paragraph we described an experiment in which a correlated error changed $S_1^1$ from 1 to $-1$, which is a large error. In order to determine if we could detect smaller errors, we performed measurements in which we'd expect $S_1^1$ to change from 1 to 0.81 (corresponding to a rotation of the detector half-wave plate by $\pi/20$). Figure 3(d)-(f) shows the results, where we measure $S_1^1 = 0.79 \pm 0.04$. The correlated error is detected with a statistical significance of 4 standard deviations. If a correlated error corresponding to rotation of the detector half-wave plate by $\pi/40$ is introduced we'd expect $S_1^1$ to change from 1 to 0.95. However, for this small change we find no statistically significant difference of $\Delta(\bar{S}) - \bar{1}$ from 0 in our measurements, so we are not able to detect this error.

Increasing the averaging time of the measurements will decrease errors due to fluctuations in the photon counting statistics. As such, we expect the ultimate sensitivity to correlated errors to be limited by repeatability and long-term drifts of the state preparation and measurement settings. Note that the numerical processing uses the measured expectation values $S_a^i$, not the settings themselves. Thus, it will be easier to detect correlated errors in matrix elements that are more sensitive to changes in the state preparation and measurement settings.

We have repeated the experiments described above by replacing the pure state $\hat{\rho}_H$ with the mixed state $\hat{\rho}_M$, and also by using $2n$ settings instead of $n+1$. With these changes we find no difference in the final results, as they look essentially the same as those shown in Fig. 3. We are able to detect a correlated error where $S_1^1$ is changed from 1 to 0.81, but do not detect the error if $S_1^1$ is changed from 1 to 0.95.

## D. Locating Correlated Errors

In the previous section we showed how we could detect correlated errors introduced into $S_1^1$, which corresponds to the first state preparation (row 1) and the first measurement setting (column 1). In this case the partial determinant differed from its expected value in the first row and column, for measurements with both $2n$ and $n+1$ settings, which agrees with theory for a correlated error at $S_1^1$. This helps us to identify the location of the correlated error with respect to the state and detector settings.

Now consider a correlated error that changes $S_2^2$. Experimental measurements and theoretical predictions for $S_2^2$ being changed from $-1$ to 1 are shown in Fig. 4, for measurements with both $2n$ and $n+1$ state and detector settings. For $2n$ settings $\Delta(\bar{S}) - \bar{1}$ differs from 0 in row 2 and column 2, while for $n+1$ settings it differs from 0 in row 1 and column 1. In both cases the difference from 0 is at least 7 standard deviations, and the experimental results are in agreement with the theoretical predications.

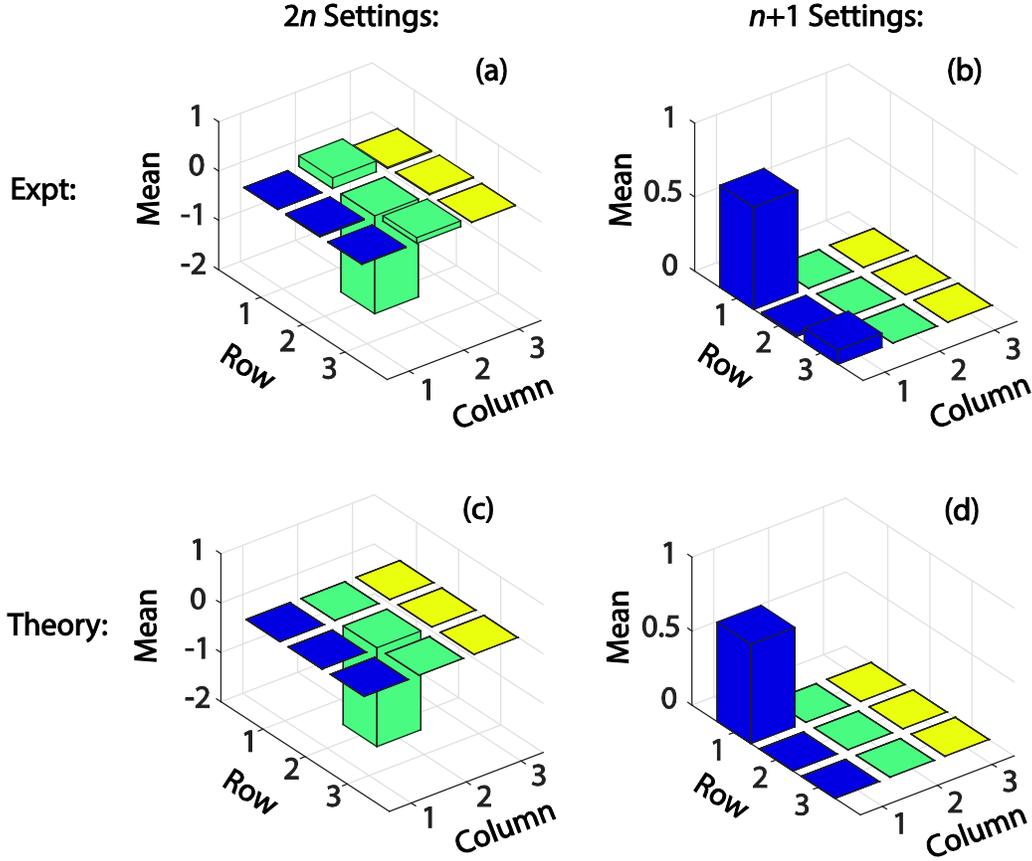

Fig. 4. For measurements with SPAM correlations in $S_2^2$ we show the mean of $\Delta(\bar{S}) - \bar{1}$ for: (a) and (b) experimental measurements, and (c) and (d) theoretical predictions. In (a) and (c) $2n$ settings are used, while in (b) and (d) $n+1$ settings are used.

By comparing Figs. 3 and 4, we see that when using $2n$ settings correlated errors in $S_1^1$ $S_2^2$ manifest themselves in different matrix elements of the partial determinant, in a manner that allows us to identify where the correlated error occurs. However, when using $n+1$ settings these correlated errors modify the same matrix element of the partial determinant; in this case we are able to identify the presence of a correlated error, but not which settings it corresponds to.

We have also performed experiments for correlated errors occurring in $S_1^2$ and $S_1^3$. For errors in these matrix elements we are able to identify the location of the correlated error when using $2n$ settings, as $\Delta(\bar{S}) - \bar{1}$ differs from 0 in the same row and column as the correlated error in $\bar{S}$. When using $n+1$ settings $\Delta(\bar{S}) - \bar{1}$ differs from 0 in the same row and column as the error in $\bar{S}$, but it *also* differs from 0 in row 1 and column 1. In all cases the experiments agree with the theoretical predictions.

When using $n+1$ settings, if the correlated error appears in row or column 2 or 3 of $\bar{S}$ (assuming $n=3$), this error is duplicated into rows and columns 5 and 6 when embedding the measured data into a $2n \times 2n$ matrix. As such, this error would appear 4 times in $\bar{S}$, whereas it only appears a single time when using $2n$ settings. This is the likely difference between the results for $2n$ and $n+1$ settings.

Furthermore, we have performed measurements with two and three different correlated errors present. We are able to identify the presence of correlated errors with both $2n$ and $n+1$ settings, and to identify the location of all of the errors when using $2n$ settings. In all cases the experiments agree with the theoretical predictions.

Since using $n+1$ settings requires less than half the number of measurements, and is able to detect correlated SPAM errors, we suggest that any experiment looking for SPAM correlations start by using $n+1$ settings to determine if there are any errors. If an error is found then measurements can be performed with $2n$ settings to better identify which state is correlated with which measurement.

## IV. CONCLUSIONS

We have performed experiments demonstrating that we can detect correlated errors between the state preparations and the measurements of a qubit using loop SPAM tomography. This determination is made by checking for self-consistency while going around a loop in parameter space. To do this one needs no knowledge about the state preparations or the measurements, other than knowing the dimensions of their Hilbert spaces. Indeed, we do not even need to estimate any state or measurement parameters in order to detect correlated errors. We find that when using $2n$ state preparations and measurement settings we are able to determine which state preparation is correlated with which measurement. When using $n+1$ settings we can determine the presence of a correlated error, but it is not clear which state preparation is correlated with which measurement. Furthermore, by having sufficient knowledge about some of the state preparations or measurement settings, we are able to use QST and QDT to estimate the rest of density operators and POVMs.

## ACKNOWLEDGEMENTS

We thank J. Spring, A. J. Menssen, W. S. Kolthammer and I. A. Walmsley for assistance with the coincidence counting unit. A.F.M. and M.B. were supported in part by the Whitman College Louis B. Perry Summer Research Endowment, the Parents Student-Faculty Research fund and the Bleakney Memorial Physics Endowment. S.J.v.E. was supported in part by ARO/LPS under Contract No. W911NF-14-C-0048.

________________